\documentstyle[12pt, epsfig]{article} \textwidth 17cm \textheight 24cm
\begin{document}

\vskip .5cm
\vbox{\vspace{6mm}}
\begin{center}
  {\large \bf \baselineskip=24pt ON SCHIZOPHRENIC EXPERIENCES OF THE
    NEUTRON \break OR \break WHY WE SHOULD BELIEVE \break IN THE
    MANY-WORLDS INTERPRETATION \break OF QUANTUM THEORY\break \break
    Lev Vaidman} \footnote[0]{School of Physics and Astronomy, Raymond
    and Beverly Sackler Faculty of Exact Sciences, Tel--Aviv
    University, Tel-Aviv 69978, Israel}

\end{center}
\vspace{2mm}

\rightline{\begin{minipage}{13.5cm}
 The truth about physical objects  must be strange.  It
may be unattainable, 
but if any philosopher believes that he has
attained it, the fact that what he 
offers as the truth is strange
ought not to be made a ground of objection to 
his opinion.\\
\rightline{-- \ \   Bertrand Russell}
\end{minipage}}

\vspace{.3cm}


\vskip .9cm

\centerline {\bf 1.  Introduction}
\vskip .5cm
 
There are many interpretations of quantum mechanics, and new ones
continue to appear.  The Many-Worlds Interpretation (MWI) introduced
by Everett (1957) impresses me as the best candidate for {\it the}
interpretation of quantum theory.  My belief is not based on a
philosophical affinity for the idea of plurality of worlds as in Lewis
(1986), but on a judgment that the physical difficulties of other
interpretations are more serious.  However, the scope of this paper
does not allow a comparative analysis of all alternatives, and my main
purpose here is to present my version of MWI, to explain why I believe
it is true, and to answer some common criticisms of MWI.

The MWI is not a theory about many {\it objective} ``worlds".  A
mathematical formalism by itself does not define the concept of a
``world".  The ``world'' is a subjective concept of a sentient
observer.  All (subjective) worlds are incorporated in {\it one}
objective Universe.  I think, however, that the name Many-Worlds
Interpretation does represent this theory fairly well.  Indeed,
according to MWI (and contrary to the standard approach) there are
{\it many worlds} of the sort we call in everyday life ``the world".
And although MWI is not just an {\it interpretation} of quantum theory
-- it differs from the standard quantum theory in certain experimental
predictions -- interpretation is an essential part of MWI; it explains
the tremendous gap between what we experience as our world and what
appears in the formalism of the quantum state of the Universe.
Schr\"odinger's equation (the basic equation of quantum theory)
predicts very accurately the results of experiments performed on
microscopic systems.  I shall argue in what follows that it also
implies the existence of many worlds.  The purpose of addition of the
collapse postulate, which represents the difference between MWI and
the standard approach,\footnote{In fact, there are several other
  interpretations without collapse.  I consider these interpretations
  to be variations of MWI. Indeed, if there is no collapse then the
  states corresponding to all worlds of MWI exist.  See more in
  Section 15.} is to escape the implications of Schr\"odinger's
equation for the existence of many worlds.

Today's technology does not allow us to test the existence of the
``other'' worlds.  So only God or ``superman'' (i.e., a
superintelligence equipped with supertechnology) can take full
advantage of MWI. We, however, are in the position of God relative to
a neutron.  Today's technology allows us to test the existence of many
``worlds'' for the neutron.  This is why I discuss neutrons first.
For the purposes of exposition I shall attribute to the neutron the
ability to feel, to remember, and to understand.  But I emphasize that
the validity of MWI held by a human observer does {\it not} depend on
the existence of a sentient neutron.

The plan of this paper is as follows: In Sections 2 and 3, I explain
the design of a neutron interferometer and show that a conscious
neutron passing through the interferometer {\it must} have
schizophrenic experiences.  In Section 4, I introduce a neutron's MWI
and explain how it solves the problem of the neutron's schizophrenia.
In Sections 5-10, I continue the discussion of MWI using the example
of the neutron interferometer.  In Section 11 I present, and in
sections 12-14 I discuss, the MWI of the Universe. Section 15 is
devoted to the {\it causal interpretation} (Bohm, 1952) which is
probably the best alternative to MWI.  In Section 16, I summarize the
arguments in favor of MWI. Section 17 is an addition to the paper in
which I reflect on recent symposium on the Many-Minds Interpretation
of Lockwood (1996).

\vskip 1cm 
\centerline{\bf 2.  The Neutron Beam Splitter} 
\vskip .2cm

        Let me start with an analysis of a simple experiment.  A
neutron passes through a beam splitter $S$ toward detectors $D_1$ and
$D_2$ (see Figure 1).  The outcome of this experiment, as reported by
numerous experimenters, is always as follows: A single neutron
coming toward the beam splitter is detected {\it either} by detector
$D_1$ {\it or} by detector $D_2$.  A natural conclusion from these
reports is that the neutron {\it either} takes trajectory $SD_1$ {\it
or} takes trajectory $SD_2$ and, consequently, the experimenter sees
only one triggered detector.  There are two distinct possibilities
and only one of them is realized. 

\epsfysize=7 cm
 \centerline{\epsfbox{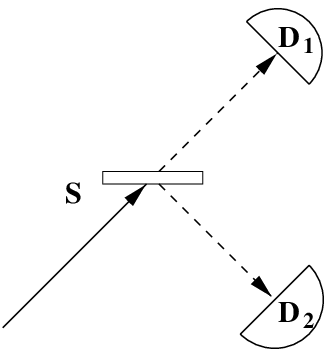}} 
{\small {\bf Figure 1}: {\footnotesize 
The neutron beam splitter.
}}
\vskip .2cm

 Before the experiment, we can imagine two different worlds
corresponding to the two possible outcomes of the experiment.  The two
worlds differ with respect to the position of the neutron, the states
of the detectors, the state of mind of the experimenter, the record
in the his notebook, etc.  In the standard approach, only
{\it one} of these worlds exists.  According to MWI, however, both
possibilities of the experiment are actualized.  Both detectors $D_1$
and $D_2$ are triggered, both outcomes are seen by the experimenter,
both results are written down in the notebook, etc.  When an
experimenter reports to me that the neutron was detected by $D_1$, I,
Lev Vaidman, {\it know} that there is also a world in which Lev
Vaidman got a report about a neutron detected by $D_2$, and that the
other world is not less ``actual'' than the first one.  This is what
``many worlds'' means.  There are many worlds like the one we
experience.

\epsfysize=17 cm
 \centerline{\epsfbox{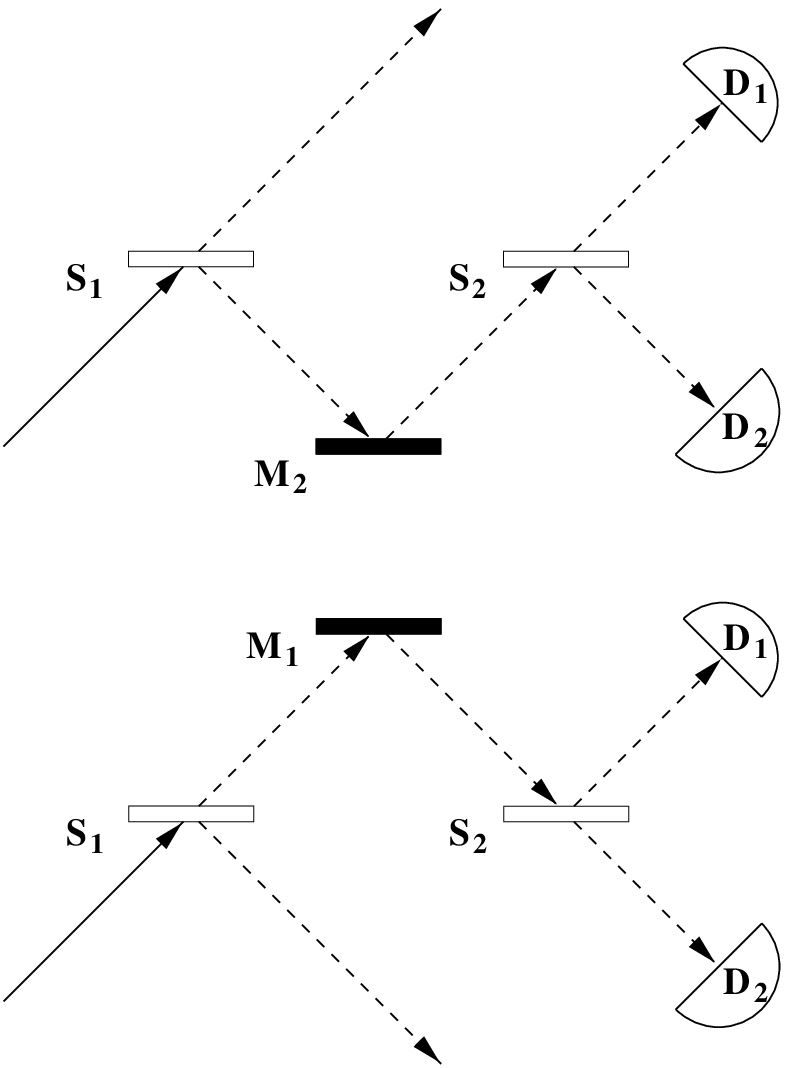}} 
{\small {\bf Figure 2}: {\footnotesize 
Two arrangements of neutron beam  splitters and a neutron mirror.
}}
\vskip .2cm

I will concede that based on the results of the experiment shown in
Figure 1, it is natural to assume that there is only one world: {\it A
neutron passing through a beam splitter either is scattered through a
given angle or continues in a straight line without being disturbed.}
The neutron has a single trajectory.  We can bolster our confidence
that this is the correct description by considering results of the
experiments with a mirror and two beam splitters in the configurations
of Figure 2.  The prediction for the outcomes of these experiments is
that in half of the trials the neutron is not detected by either of
the two detectors (when it takes the trajectory without the mirror),
and in the other half it is detected at random by $D_1$ and $D_2$.
The experimental results are, indeed, as predicted.  However, when we
combine these two systems, we discover that what was true for each of
the systems individually is not true anymore: the neutrons are not
detected at random by $D_1$ and $D_2$.  This combination of two beam
splitters and two mirrors is called a {\it neutron interferometer} and
I will discuss it in the next section.

\vskip 1cm

   \centerline{\bf 3.  The Neutron Interferometer}
\vskip .2cm

The neutron interferometer is an experimental device that can be found
in several laboratories in the world, for a comprehensive review see
Greenberger (1983).  Taking the assumption of the previous section it
is impossible to explain the results of the neutron interference
experiment.  These results, combined with the assumption that there is
only one world for the neutron, compel the neutron to have
schizophrenic experiences.

In Figure 3, a schematic neutron interference experimental setup is
shown.  It consists of a source of neutrons, a beam splitter $S_1$,
two mirrors $M_1$ and $M_2$, another beam splitter $S_2$, and two
detectors $D_1$ and $D_2$.  Based on our understanding of the process
of a neutron passing through a beam splitter, i.e., that it either is
scattered through a given angle or continues in a straight line
without being disturbed, we conclude that each neutron takes one of
the four trajectories $S_1M_1S_2D_1$, $S_1M_1S_2D_2$, $S_1M_2S_2D_1$,
$S_1M_2S_2D_2$.  Therefore, the neutrons have to be detected at random
by detectors $D_1$ and $D_2$.  But the experiment does not show what
is expected! {\it All} the neutrons are detected by detector
$D_1$.\footnote{Here (and in a few places below) I sacrifice rigor for
  simplicity by omitting technical details.  A precise statement here
  is that one {\it can tune} the interferometer such that all neutrons
  are detected by $D_1$.}

\epsfysize=8 cm \centerline{\epsfbox{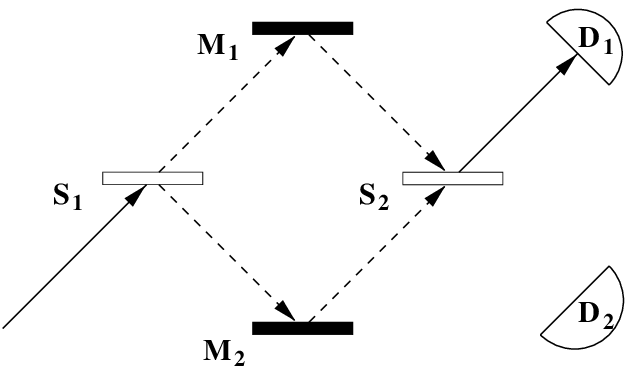}} {\small {\bf Figure 3}:
  {\footnotesize The neutron interferometer.  }} \vskip .2cm

        We cannot explain the experimental results by the picture of a
single trajectory for the neutron.  We are compelled to admit that in
some sense the single neutron passes through {\it two} separate
trajectories: $S_1M_1S_2$ {\it and} $S_1M_2S_2$.  If the neutron can feel, it
experiences being in two places and moving in two different directions
simultaneously.  Inside the interferometer the neutron must therefore
have schizophrenic experiences.\footnote{ The word ``schizophrenia'' does not describe precisely
the neutron's experience, but I cannot find a better alternative.  The
difficulty in language is not surprising since, before quantum
mechanics, humans had no reason to discuss this kind of situation.}

\vskip 1cm

 \centerline{\bf 4.  Two Neutron Worlds} 
   \vskip .2cm

        To avoid positing schizophrenic neutrons, I will state that
during the time the neutron is inside the interferometer the world of
the experimenter encompasses {\it two} neutron worlds.  In each of
these two worlds, the neutron has a definite trajectory: $S_1M_1S_2$
for one and $S_1M_2S_2$ for the other.  In each world there is a
causal chain of events.  For example, in one world the neutron passed
through beam splitter $S_1$ undisturbed,  kicked by mirror $M_1$ bounced
toward $S_2$, was scattered by the beam splitter toward detector $D_1$, and
was absorbed by $D_1$.  In each world the neutron has unambiguous answers
to the questions: Where is the neutron now?  What is the direction of
its motion?  Which mirror did it hit?  Note that my assumption of two
neutron worlds is useful even if there are no sentient neutrons.  The
assumption allows {\it me} to answer the above questions, questions
which are illegitimate according to the standard approach. 

The neutron in one neutron world does not know (unless it has studied
quantum mechanics and believes in MWI) about the existence of its
``twin'' in the other world.  In the same way, most of us do not think
that in addition to the world we experience there are other worlds
present in the space-time.  The experimenter, however, is in the
position of God for the neutron.  He can devise an experiment to test
whether the neutron of one world feels the neutron of the other world.
To this end he modifies the neutron interference experiment by
removing beam splitter $S_2$, see Figure 4.  One neutron world
corresponds to the trajectory $S_1M_1D_2$ and the other to the
trajectory $S_1M_2D_1$.  We know that the two neutrons meet each other
in at point $A$, the original location of beam splitter $S_2$.  They
are in the same place at the same time moving in different directions.
Under normal circumstances (in a single world) two neutrons would
scatter from each other.  But the result of the experiment, as in
Figure 4, shows no scattering whatsoever.  The rate of detection of
neutrons by $D_1$ ($D_2$) is {\it not} affected in any way when we
eliminate the twin-neutrons by placing an absorption screen before
mirror $M_1$ ($M_2$).

\epsfysize=8.5 cm
 \centerline{\epsfbox{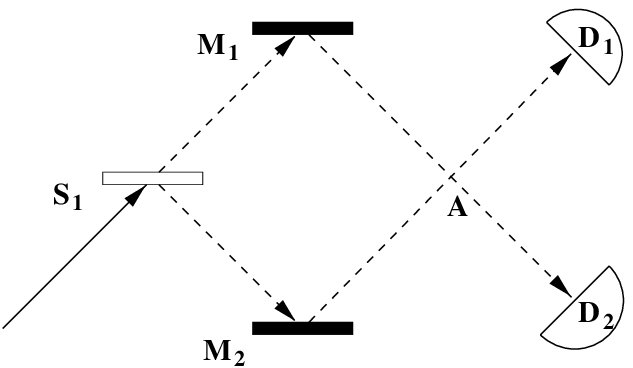}} 
{\small {\bf Figure 4}: {\footnotesize 
The neutron interferometer without second beam  splitter.
}}
\vskip .2cm

Let us discuss again the neutron interference experiment (Figure 3).
The hypothesis of many (two in this case) worlds solves the problem of
the neutron's schizophrenia inside the interferometer, but it seems
that we are left with the problem of schizophrenic {\it memories} of
the neutron.  The two worlds become one again in beam splitter $S_2$.
What memory does the neutron have after it leaves $S_2$?  Did it hit
$M_1$ or $M_2$?  Quantum theory tells us that the neutron {\it cannot}
retain memories about which trajectory it took (in which world it
``lived").  Quantum mechanics does explain why the neutron is detected
by $D_1$, but only if the neutron has no internal variable that
``remembers'' (after the neutron leaves the interferometer) which
trajectory the neutron has taken.  Neutron memory is not ruled out
completely: the neutron might remember its trajectory while it is
still inside the interferometer, but the memory has to be erased when
the neutron leaves the second beam splitter.  In fact, there is a
physical realization of an experiment in which the neutron
``remembers", while inside the interferometer, which path it takes.
One of the devices that can serve as a beam splitter for a neutron is
a specially designed magnet (Stern-Gerlach apparatus).  In this case,
the path of the neutron is correlated with the value of an internal
variable called spin, so the neutron has the spin to remind itself of
its path while it is inside the interferometer.  However, the second
magnet, replacing the second beam splitter, erases the correlation and
the memory once the neutron leaves the interferometer.

The neutron cannot ``feel'' objects from other worlds, it cannot
remember that it ``lived'' in two worlds.  So, is there any reason for
the neutron to believe in the existence of the other worlds?  Yes, the
same reason that we have: this hypothesis explains why, after passing
through the interferometer, the neutron always ends up in detector
$D_1$.  We will see next, how the quantum theory (with many worlds)
explains this experimental fact.

\vskip 1cm

 \centerline{\bf 5.  Quantum-Mechanical Explanation}
\vskip .2cm
 
In standard quantum mechanics particles do not and {\it cannot} have
trajectories.  A particle is described by a quantum state evolving in
time.  For the neutron, the quantum state is represented by a spin
component and a spatial wave function.  According to the standard
interpretation, the square of the magnitude of the wave function at a
given point yields the probability per unit volume of finding the
particle there.  Frequently, the spatial wave function spreads out
significantly and then there is no answer to the question: where is
the particle?  In fact, in any real situation there is no exact answer
to this question.  (Zero uncertainty in the position requires infinite
energy.)  Nevertheless, physicists do consider trajectories of
particles.  What physicists mean when they say that the neutron takes
a given trajectory is that the spatial wave function of the neutron is
a {\it localized wave packet} (LWP) whose center moves on this
trajectory. (Macroscopic bodies are also described by LWPs.  Even a
leading theory of collapse, see Ghirardi and Pearle (1990), considers
reduction of the quantum states of macroscopic bodies only into LWPs.)
Inside the interferometer the wave function of the neutron is not a
LWP and, consequently, the neutron has no trajectory.  However, when
the neutron leaves the beam splitter $S_2$, its wave function becomes
again a LWP, the LWP which moves toward detector $D_1$.  This is the
quantum-mechanical explanation why the neutron is never detected by
detector $D_2$.  Let me demonstrate now this quantum interference
effect using some formulas.

I designate by $|up\rangle$ and $|down\rangle$ the states of the
neutron moving 45$^0$ up and 45$^0$ down respectively (see Figures 1-4).
After the passage through a beam splitter the state of the neutron
changes as follows:
$$
|up\rangle \rightarrow 1/\sqrt 2  (|up\rangle + |down\rangle) ,
$$
$$\eqno(1)$$
$$
|down\rangle \rightarrow 1/\sqrt 2  (|up\rangle - |down\rangle) .
$$
The action of mirror $M_1$ is
$$
|up\rangle \rightarrow |down\rangle, \eqno(2)
$$
and of mirror $M_2$,
$$
|down\rangle \rightarrow |up\rangle. \eqno(3)
$$
Knowing the action of components (1) - (3), and using the linearity of
quantum mechanics, we can find out the state of the neutron leaving
the interferometer:

$$
|up\rangle \rightarrow 1/\sqrt 2 (|up\rangle + |down\rangle)
\rightarrow 1/\sqrt 2 (|down\rangle + |up\rangle)
~~~~~~~~~~~~~~~~~~~~~~~$$
$$ ~~~~~~~~~~~~~~~~~~~~~~\rightarrow  1/ 2  (|up\rangle - |down\rangle)
+ 1/ 2  (|up\rangle + |down\rangle) = |up \rangle \eqno(4)
$$
 The neutron LWP, after leaving the beam splitter $S_2$, moves in the
direction ``up'' and is absorbed by detector $D_1$.  This explanation is so
simple that it is generally accepted even though it involves an
intermediate state of the neutron moving {\it both} up and down at
the same time.

We can also understand why the neutron interference experiment cannot
be explained if the neutron remembers which path it took.  If it has a
memory variable $M_i$ corresponding to which mirror it hit, the two
waves reaching detector $D_2$ are different and, therefore, do not
interfere.  The corresponding terms in the state of the neutron, $
-1/2 |down, M_1 \rangle$ and $+ 1/2 |down, M_2 \rangle$ are not
canceled as are the terms $ -1/2 |down \rangle$ and $+ 1/2 |down
\rangle$ in Eq.  (4).

The neutron inside the interferometer is described by the wave
function that is a superposition of two LWPs distinguished by their
direction of motion and location:
$$
|\Psi \rangle_{\rm neutron} = 1/\sqrt 2  (|up\rangle + |down\rangle) .\eqno(5)
$$
In the standard approach, a sentient neutron would invariably be
schizophrenic.  My proposal is that during the period of time the
neutron wave function is inside the interferometer {\it there are two
  neutron worlds} : one corresponding to LWP $|up\rangle$ and the
other to LWP $|down\rangle$.  In each world there is a neutron with
its own trajectory.  We can view a part of the neutron wave function
as a ``whole'' neutron (in a given world) because physical
characteristics of the ``partial" neutron such as mass, spin, etc. are
exactly the same as the characteristics of the whole neutron.  The
trajectory of each LWP (inside the interferometer where there are no
splittings) is just what it would be if it were the whole wave
function. The neutron in each world cannot know from immediate
experience that in some sense it is only ``half'' a neutron. Indeed,
any physical measurements performed by the ``half'' neutron moving in
one arm of the interferometer would yield exactly the same results as
the same measurements performed by the ``whole'' neutron moving in
this arm.

\vskip 1cm

\centerline{\bf 6.  The Preferred Basis of the Neutron Worlds}

   \vskip .2cm

   In the previous section I decomposed the quantum state of the
   neutron (5) into a sum of two orthogonal states corresponding to
   two different neutron worlds.  In the formalism of quantum
   mechanics there are infinitely many ways to decompose the state
   into a sum of two orthogonal states.  Why did I chose this
   particular one?  Why not, for example, take an alternative
   decomposition of the same state:
   $$
   |\Psi \rangle_{\rm neutron} = 1/\sqrt 8 \Bigl((1 + i) |up\rangle
   + (1-i)|down\rangle\Bigr) + 1/\sqrt 8 \Bigl((1 - i) |up\rangle +
   (1+i)|down\rangle\Bigr).\eqno(6)
$$
But the two components in Eq.  (6) do not correspond to ``neutron
worlds".  Indeed, I have made an assumption that neutrons are similar
to us, i.e., a sentient neutron is not schizophrenic as it would have
to be in the worlds corresponding to $1/\sqrt 8 ((1 + i) |up\rangle +
(1-i)|down\rangle)$ and $ 1/\sqrt 8 ((1 - i) |up\rangle +
(1+i)|down\rangle) $.  The decomposition (5) is, essentially, the only
decomposition into ``worlds'' in which the neutron is a localized wave
packet during the whole period of time and, therefore, has a single
experience at every moment.  It is possible to decompose each term in
Eq.  (5) into smaller LWPs and if the neutron can distinguish between
the trajectories of these LWPs, the decomposition should be made into
more than two neutron worlds.

I want to stress that ``world'' is not a physical concept. It is
defined by the conscious mind of the observer.  ``Physics'' does not
prefer the decomposition into LWPs. It is the fact that the observer
has local senses that explains why an evolution of a LWP corresponds
to a definite chain of events which he perceives, a story which
defines a particular world. See more in Sections 13 and 17.
 
\vskip 1cm

 \centerline{\bf 7.  The Concept of Probability of a Believer in MWI}
    \vskip .2cm

Let me now discuss the experience of the neutron as it passes through
a beam splitter.  This is the process in which one neutron world
transforms into two worlds.  The neutron experiences one of two
possibilities: either it scatters or it remains undisturbed.
Assuming that the neutron does not know MWI, it has no reason to
believe that the other possibility is also realized.  The neutron
which passes through many beam splitters develops a concept of {\it
probability}.  The situation for the sentient neutron is the same as
for an experimenter observing the result of the experiment of Figure
1.  The neutron finds itself in detector $D_1$ or detector $D_2$, and
the experimenter finds accordingly that detector $D_1$ or $D_2$
clicks.  Thus, we can identify the experimenter's concept of probability
with the neutron's concept of probability.  The neutron passing
through the beam splitter described above in Eq.  (1) assigns equal
probabilities to the states $|up\rangle$ and $|down\rangle$.  Some
other beam splitters do not give equal probability for the two
possible results.  The general form of the operation of a beam
splitter is
$$
|up\rangle \rightarrow \alpha |up\rangle + \beta |down\rangle ,\eqno(7)
$$
Quantum theory yields for the neutron, in this case, the probability
$|\alpha|^2$ to be found in $D_1$ and the probability $|\beta|^2$ to
be found in $D_2$.

It is more difficult to define a concept of probability for those
experimenters and those neutrons who know MWI. They understand that
the belief of the neutron (it might be more correct to say ``the
belief of both neutrons"), that there is just one world, is an
illusion.  There are two worlds in parallel: one with the neutron in
the state $|up\rangle$ and the other with the neutron in the state
$|down\rangle$.  Thus, the phrase ``the probability for the neutron to
be found at $D_1$'' seems senseless.  Indeed, it is not clear what
``the neutron'' in this phrase means, and it seems that whatever
neutron we consider, we cannot obtain $|\alpha|^2$ for the
probability. For the neutron passing through a beam splitter the
probability to end up at $D_1$ {\it as opposed} to $D_2$ is
meaningless because this neutron becomes {\it two} neutrons.  The two
new neutrons are identified with the old one: the neutron detected by
$D_1$ and the neutron detected by $D_2$ {\it both} entered the beam
splitter.  The new neutrons have no identity problem; the neutron at
$D_1$ has the direct experience of being at $D_1$ as opposed to $D_2$,
but it seems that the probability for that neutron to be at $D_1$ is
just 1.  {\it We} cannot assign any other number to this probability,
but the neutron {\it can}.  Suppose that the neutron (not enjoying
beam splitters) took a sleeping pill and slept until it reached a
detector.  Now, if it awakes inside the detector but has not yet
opened its eyes, the neutron (an expert in quantum mechanics) can say:
``I have a probability $|\alpha|^2$ to find myself in $D_1$".  This is
an ``ignorance-type'' probability. We, like any external system,
cannot be ignorant about the location of the neutron since we identify
it using its location, while each sentient neutron does not need
information to identify itself.\footnote{Albert (1987) pointed out
  another interesting ``privilege'' of an observer in comparison with
  external systems: the observer is the {\it only} one in a position
  to know certain facts about himself.} The second new neutron, the
one at $D_2$, before opening his eyes has exactly the same belief: ``I
have a probability $|\alpha|^2$ to find myself in $D_1$". The neutron
entering the beam-splitter converts into two neutrons which have the
same belief about probability. This allows us to associate the
probability for the neutron entering the beam-splitter to end up at
$D_1$ as the probability of its ancestors to end up there.

The gedanken story with a ``sleeping pill'' explains how the concept
of probability can be introduced in the framework of MWI. An
experimenter preparing a quantum experiment with several possible
outcomes can associate probability for different outcomes according to
the ignorance probability of each of his ancestors to obtain this
outcome. And the sleeping pill is hardly necessary since in typical
experiment a superposition of macroscopically different states arises
before the observer(s) become aware of the result of the experiment.

 
\vskip 1cm

 \centerline{\bf 8.  The Measure of Existence of a World} 
   \vskip .2cm 
   
   A believer in MWI can define a {\it measure of existence of a
     world}, the concept which yields his subjective notion of
   probability.  The measure of existence of a world is the square of
   the magnitude of the coefficient of this world in the decomposition
   of the state of the Universe into the sum of orthogonal states
   (worlds). {\it The probability postulate} of MWI is: If a world
   with a measure $\mu$ splits into several worlds then the
   probability (in the sense above) for a sentient being to find
   itself in a world with measure $\mu_i$ (one of these several
   worlds) is equal to $\mu_i/\mu$.  See Lockwood (1989, pp. 230-232)
   for a pictorial explanation of this rule.  Consider, for example, a
   world with measure of existence $\mu$, in which a neutron enters
   the beam splitter shown in Figure 1.  Assume that the operation of
   the beam splitter is described by equation (7).  Then the measure
   of existence of the world in which the neutron reaches detector
   $D_1$ equals $\mu |\alpha|^2$, and therefore the probability for
   the neutron to find itself in $D_1$ is $\mu |\alpha|^2/ \mu =
   |\alpha|^2$.

During the time a neutron evolves as a single LWP, its measure of
existence has no physical manifestation.  All physical parameters,
such as mass, spin, magnetic moment etc., are independent of the
measure of existence.  A neutron with a tiny measure of existence
moves (feels) exactly as one with measure 1. The measure of existence
manifests itself only in processes in which splitting of the world
takes place (in the standard interpretation it corresponds to the
situations in which a collapse occurs).  The relative measures of
existence of the worlds into which the world splits provides a concept
of probability.

I believe that the argument above, explaining how the measure of existence
of future worlds yields a probability concept, is  enough to justify 
introducing the concept of ``measure of existence''. However, 
even the measure of existence of present worlds has physical meaning. 
  What is the
``advantage'' of being in a world with large measure of existence?  When
the neutron (i.e., LWP) evolves without splitting, the other worlds
cannot interfere.  When it splits into two in a beam splitter, the
other worlds usually do not interfere either, but they can!  Consider
the neutron moving inside the upper arm of the interferometer (Figure
3 ) and assume that its measure of existence equals 1/2.  Being
unaware of its ``twin'' in the bottom arm, it calculates equal
probabilities for reaching detectors $D_1$ and $D_2$.  But, the
neutron's god, namely the experimenter, makes use of the other neutron
world and {\it changes} the probabilities completely.  If, however,
the neutron in the upper arm has measure of existence $\mu \simeq 1$
(if the beam splitter $S_1$ is replaced by the one which transmits
most of the wave), then nobody, not even god, can significantly change
the quantum probabilities.  When the measure of existence is less than
or equal to 1/2, the god can change probabilities of further splitting
completely; when it is greater than 1/2, only partially, and when it
is equal to 1 the god cannot change the probabilities at all.  Even
for neutrons, experimentalists have to work hard to change such
probabilities.  A similar experiment involving human beings  (I
discuss it in Section 11) would be
astronomically difficult.  We have no indication that any god
(superintelligence from another planet) plays such a game with us.
 
\vskip 1cm

 \centerline{\bf 9.  The Collapse Postulate and Why We Do Not Need It} 

   \vskip .2cm 

        What I have done so far may be called the many (two)
neutron-worlds {\it interpretation} of a neutron interference
experiment.  I have introduced unusual language, but with regard to
equations and results of experiments, I am in complete agreement with
the standard approach.  However, the Many-Worlds Interpretation of
quantum mechanics, in spite of its name, is a {\it different
theory}.  The standard approach to quantum mechanics includes all
axioms of MWI and it has one more: the postulate of the {\it collapse}
of a quantum state in the measurement process.  The collapse postulate
has physical consequences which in principle can be tested, although
today's technology is very far from permitting a decisive experiment.

        Collapse occurs when a measurement is performed.  There is no
collapse of the neutron state inside the interferometer, and so my
discussion agrees with the standard approach.  In order to display the
differences between MWI and the standard approach let us consider a
neutron passing through a beam splitter with action described by
equation (7) and detected by detectors $D_1$ and $D_2$ (Figure 1).
According to MWI, the description of this process is:
$$
|up\rangle |r\rangle_{D_1} |r\rangle_{D_2} \rightarrow (\alpha
|up\rangle + \beta |down\rangle) |r\rangle_{D_1} |r\rangle_{D_2}
\rightarrow ~~~~~~~~~~~$$
$$~~~~~~~~~~~~~~~~ \alpha |in~D_1\rangle |in\rangle_{D_1}
|r\rangle_{D_2} + \beta |in~D_2\rangle |r\rangle_{D_1}
|in\rangle_{D_2}, \eqno(8)
$$
where $|r\rangle_{D_1}$ signifies the ``ready'' state of detector $D_1$,
$|in~D_1\rangle$ signifies the state of the neutron when it absorbed
by detector $D_1$, $|in\rangle_{D_1}$ signifies the state of detector
$D_1$ ``neutron in the detector,'' etc.  Because of the collapse
postulate, the final state (8) immediately transforms (with the
appropriate probability) into a state with a definite result of the
experiment:
$$
\alpha |in~D_1\rangle |in\rangle_{D_1} |r\rangle_{D_2} + \beta |in
D_2\rangle |r\rangle_{D_1} |in\rangle_{D_2}\rightarrow
~~~~~~~~~~~~~~~~~~~~~~~$$
$$~~~~~~~~~~~~~~~ \rightarrow 
\cases{|in~D_1\rangle |in\rangle_{D_1} |r\rangle_{D_2}
~~~~~~~~~~~&(probability $|\alpha|^2$) ~~ or\cr
|in~D_2\rangle |r\rangle_{D_1} |in\rangle_{D_2}
~~~~~~~~~~~&(probability $|\beta|^2)$\cr} \eqno(9)
$$
 The motivation for this step is obvious.  The right hand side of (8)
indicates that at the end of the measurement detector $D_1$ registers
`in' {\it and} detector $D_2$ registers `in' (as well as that both
detectors show `{\it r}').  The experimenters, however, always report
that a {\it single} detector registers ``in".

        It seems that the collapse postulate is necessary to explain
the experimental results.  This, however, is not the case.  Quantum
mechanics without the collapse postulate explains the reports of the
experimenters as well.  Indeed, let me also consider the experimenter
as a quantum system.  Quantum mechanics describes the process of
observation (when the state of the neutron and the detectors are
described by equation (8)) as follows:
$$
\Bigl(\alpha |in~D_1\rangle |in\rangle_{D_1} |r\rangle_{D_2} +
\beta |in~D_2\rangle |r\rangle_{D_1}
|in\rangle_{D_2}\Bigr)~|r\rangle_E \rightarrow
~~~~~~~~~~~~~~~~~~~~~~~~~~~~
$$
$$ \rightarrow
\alpha |in~D_1\rangle |in\rangle_{D_1} |r\rangle_{D_2}
|see~D_1~ 'in',~D_2~ 'r'\rangle_E ~~~~~~~~~~~~~~~~~~~~~~~~~$$
$$~~~~~~~~~~~~~~~~~~~~~~~ +~ \beta |in~D_2\rangle
|r\rangle_{D_1} |in\rangle_{D_2})|see~D_1~ 'r',~ D_2~ 'in'\rangle_E
\eqno(10)
$$
where $|see~ D_1~ 'in',~D_2~ 'r'\rangle_E$ signifies the state of
the experimenter seeing $D_1$ clicks, $D_2$ ``ready", etc.  In quantum
mechanics without collapse, there is no experimenter who sees the
neutron being detected by both detectors.  Instead, there are two
different experimenters: one reports that the neutron is detected by
$D_1$ and is not detected by $D_2$, and the other reports that the
neutron is detected by $D_2$ and is not detected by $D_1$.  Why are we
never confused by their contradictory reports?  Because we, in turn,
by listening to their reports, are also splitting in the same way.
And any other experimenter who observes the detectors splits.  After
the experiment there are two worlds: in one of them all agree that the
neutron is in $D_1$, and in the other all agree that the neutron is in
$D_2$.  Both worlds are real.  If I got a report that the neutron is
in $D_1$, I should not believe that this world is more real than the
world in which the neutron is in $D_2$.  It might be that their
measures of existence are different, i.e., in some sense, there is
``more'' of one world than of the other.  However, I still should not
say that this world is more real than the other.  There is no reason
whatsoever to believe that the measure of existence of the world in
which you now read this paper is maximal among all worlds, but
nevertheless it is as real as it can be.
 
 \vskip 1cm
 
 \centerline{\bf 10.  Test of MWI} 
\vskip .2cm
    
A widespread misconception about MWI is that its predictions are
identical to the predictions of the standard approach (e.g. De Witt,
1970).  Let me describe here the design of an experiment that
distinguishes between MWI and standard (collapse) approach (see also
Deutsch (1986) and Lockwood (1989, p.223).  The measurement process of
the experiment illustrated in Figure 1, including the observation of
its result by an experimenter, can be described (if MWI is a correct
theory) by Schr\"odinger's equation with a certain Hamiltonian.  A
``superman'' could build a device with a ``time reversal'' Hamiltonian
which could ``undo'' the measurement.  The ``time reversal''
Hamiltonian would erase the memory of the experimenter, the detectors
would return to the ``ready'' state, and the neutron would return to
its original place, i.e. the neutron's source.  At this stage we
replace the source of the neutron by a detector.  If no collapse takes
place, the detector will detect the neutron with probability 1. The
neutron in its ``reverse'' motion arrives at the beam splitter from
two directions and, as in the neutron interference experiment (Figure
3), continues in a single direction toward the detector.  If, however,
the collapse takes place at some stage during the measuring procedure
-- say, when the experimenter looks at the detectors -- then the
neutron in its ``reverse'' motion arrives at the beam splitter only
from one direction.  Consequently, it comes out of the beam splitter
in {\it two} directions (see Eq.  (1)).  In this case the probability
of detecting the neutron is equal to 1/2.  Thus, MWI will be confirmed
if the neutron is always detected by the detector, and it will be
refuted if the neutron is detected in only about half of the trials.

Since it is generally believed that the collapse happens when the
neutron is detected by a macroscopic detector, an experiment which
does not involve a human observer is also a reasonable test of MWI.
If the detector is microscopic, then it is even feasible now to design
the device which undoes the interaction between the neutron and the
detectors.  With progress in technology, we can get closer and closer
to a decisive experiment.  A new experimental field, two-particle
interferometry (Horne, Shimony and Zeilinger 1989), is a significant
step toward this goal.  While in the case of a neutron interferometer,
the two worlds which were made differed only with respect to the
trajectory of a {\it single} neutron, now the worlds which interfere
with each other differ with respect to the trajectories of {\it two}
particles.
 
 \vskip 1cm

 \centerline{\bf 11.  MWI as a Universal Theory} 

   \vskip .2cm  
   
   According to MWI the Universe, everything that exists, is
   characterized by a single quantum state, the State.  The time
   evolution of the State is completely deterministic (given by
   Schr\"odinger's equation).  Essentially, the Universe {\it is} the
   State.  The world, as we commonly understand it through our
   experience, corresponds to a tiny part of this State, and we, to
   some fragment of this part. I see remote support for this picture
   in recent work of Redhead and Teller (1992) denying individuability
   of identical particles.  Thus, in particular, we are not to label
   the electrons, the protons, etc. out of which {\it we} are made.
   What specifies and defines {\it us} is the configuration, the shape
   of the fragment of the state corresponding to our world.

 The State $|\Psi\rangle$ can be decomposed
into a superposition of orthogonal states $|\psi_i\rangle$
corresponding to different worlds:
$$
 |\Psi\rangle = \sum_i \alpha_i |\psi_i\rangle \eqno(11)
 $$
 The basis of the decomposition (11) of the Universe is determined
 by the requirement that individual terms $|\psi_i\rangle$ correspond
 to sensible worlds.  The consciousness of sentient beings who are
 attempting to describe the Universe {\it defines} this basis.  I want
 to emphasize that the choice of the basis has no effect whatsoever on
 the time evolution of the Universe.  The concept of {\it world} in
 MWI is not part of the mathematical theory, but a subjective entity
 connected to the perception of the observer (e.g. sentient neutron),
 such that it corresponds for human beings to our usual notion of the
 world.  In this context one can understand speculation of Wigner
 (1962) about the collapse caused by the consciousness of the
 observer, but not in a literal sense, i.e., that there is a law
 according to which consciousness affects physical processes.
 Instead, the conscious observer defines the basis of decomposition of
 the Universe into the worlds.  Thus, one experimenter's world
 encompasses two (sentient) neutron worlds.  I analyze this ``observer
 decomposition'' in Section 13 (see also Ben-Dov, 1990).
 
The coefficients of the equation (11) yield measures of existence of
different worlds. The measure of existence of the world
$|\psi_i\rangle$ is $|\alpha_i|^2$.  Although we do not experience it
directly, I can, as above, discuss two manifestations of the measure
of existence.  The first manifestation is for the future worlds.
Every time there is a situation in which the world splits it is
important for a believer in MWI to know the relative measures of
existence of the splitted worlds.  If asked, he will bet according to
these numbers. In particular, for the experiment described in Figure 1
in which the neutron passes through, say $10\%-90\%$ beam splitter he
will bet 1:9 for the neutron reaching corresponding detectors. He
understands that he has an illusion of corresponding probabilities
even so no random process take place in the Universe. In fact, this
behavior of the believer in MWI will be identical to a (normal)
behavior of a believer in the collapse governed by Born probability
rule.
The second manifestation, which can be seen only in a gedanken
experiment, is for the measures of existence of the present worlds.  I
will show that in a certain situation we should behave differently
just because of the different values of the measure of existence of
corresponding worlds.

Let us assume that tomorrow a ``superman'' will land on Earth. He is
far more advanced in technology than we are, and he will show to us
that he can perform interference experiments with macroscopic bodies.
He will resurrect Schr\"odinger cats, ``undo'' measurements described
in Section 10 (showing that no collapse takes place and that MWI is
correct) etc.  He also will convince us that we can rely on his word.
Then he will offer me a bet, say 1:1, that the neutron which passes
through a $10\%-90\%$ beam splitter as above will end up in detector
$D_1$ (corresponding to $10\%$ probability calculated naively). He
will promise not to touch {\it this} neutron, i.e., the neutron coming
$45^0$ up. Now it is important for my decision of accepting or
rejecting the bet to know my {\it present} measure of existence. I
remember that after the superman's landing I performed a quantum
experiment and obtained a very improbable result. This means, that the
measure of existence of my world is very small relative to that in
which there is Lev Vaidman to whom the superman with his
super-technology also has an access. Thus, the superman can, in
principle, change the state of my twin in this other world (including
the twin's memory) making it identical to that of mine and send in the
other world the neutron $45^0$ from the top, arranging, via
interference of the two worlds, zero probability for the detection by
detector $D_2$ (which had $90\%$ probability without the actions of
the superman). So, in that case I {\it should not} take the bet. If,
however, I know that I have a large measure of existence compare to
twins with which the superman might play, then I {\it should} take the
bet, since the superman, in spite of his unlimited technological
power, cannot change significantly the probabilities of the
measurement outcomes (the measures of existence of corresponding
worlds) .

 \vskip 1cm

 \centerline{\bf 12.  How Many Worlds?}
    \vskip .2cm 
    
    Healey (1984) and many others became opponents of MWI trying to
    answer the question of how many worlds there are.  The number of
    worlds is huge, and it is not clear how to define it rigorously.
    Nevertheless, I do not see this as a serious problem, because the
    number of worlds {\it is not} a physical parameter in the theory.
    The physical theory is about the Universe, {\it one} Universe.
    Worlds are {\it subjective} concepts of the observers.  A world is
    a sensible description.  It can be characterized by the values of
    a set of variables.  If the State (of the Universe) is known, one
    can calculate the expectation value of a projection operator
    corresponding to these values of the set of variables.  It is
    equal to the measure of existence of this world.  If the measure
    is zero, I define that the world does not exist.  I do not know
    the State.  Therefore, I do not know if any particular world
    exists.  I do know that the world in which I wrote this paper
    exists.  I also have knowledge about quantum experiments with
    possible different outcomes which were performed in the past.
    Therefore, I know that there are other worlds.  And the worlds
    continue to multiply.  By performing quantum experiments with {\it
      a priori} uncertain outcomes, I am certain that I increase the
    number of worlds.  (I disregard improbable situations in which the
    worlds recombine.) I tend to believe that even without special
    designs of quantum-type experiments, there are numerous processes
    which split the worlds.  This question can be resolved by careful
    analysis using the standard approach.  Every time we encounter a
    situation in which, in the standard approach, collapse must take
    place, splitting takes place; and the ambiguity connected with the
    stage at which collapse occurs corresponds to the subjective
    nature of the concept of world.\footnote{ While this ambiguity
      represents a very serious conceptual difficulty of the collapse
      theories, it is not a serious problem in the MWI. The collapse
      as a physical process should not be vaguely defined, while the
      framework of concepts of conscious beings might have a lot of
      freedom.} There are very many worlds from the perspective of
    human beings, although not as many as in the modal realism
    approach of Lewis (1986) in which {\it every} logically possible
    world exists.  See a comparative analysis by Skyrms (1976).

 \vskip 1cm

 \centerline{\bf 13.  Locality of the Preferred Basis}

    \vskip .2cm 
    
    Let me sketch a conjecture about a theory of evolution of sentient
    observers with local senses such as we possess.  Consciousness is
    a collection of thoughts.  Thoughts are representations of causal
    chains of events.  Events are describable in terms of observer's
    experiences.  The experiences are obtained through the senses in a
    process explainable by physical interactions.  Physical
    interactions are local.  These are the reasons why causal chains
    represented by our thoughts consist of {\it local} events.  The
    neutron, ``created'' here ``in the human image'', can understand
    local events such as hitting a mirror, while it cannot comprehend
    the experience of being in two places simultaneously.  The neutron
    distinguishes between local worlds given by Eq.  (5) and cannot
    distinguish among orthogonal nonlocal states as in Eq.  (6).

Physics explains why an observer who ``thinks'' in the concepts of
nonlocal superpositions is not favored by evolution.  Imagine an
observer who can distinguish between two nonlocal orthogonal states of
a macroscopic system.  He ``thinks'' in the concepts of nonlocal
superpositions and acts differently according to orthogonal nonlocal
states.  For example, if the state of the neutron and the detectors in
the experiment of Figure 1 is
$$
1/\sqrt 2 (|in~D_1\rangle |in\rangle_{D_1} |r\rangle_{D_2} +
|in~D_2\rangle |r\rangle_{D_1} |in\rangle_{D_2}) , \eqno(12a)
 $$
 he makes a record ``+'' in his notebook, and if the state is $$
 1/\sqrt 2 (|in~D_1\rangle |in\rangle_{D_1} |r\rangle_{D_2} -
 |in~D_2\rangle |r\rangle_{D_1} |in\rangle_{D_2}) , \eqno(12a)
$$
he makes a record ``$-$".  However, these records will not be helpful
because through local interactions with the environment the system
consisting of the neutron and two detectors will in both cases soon
cease to be in a pure quantum state.  The system will be described by
a mixture with equal probability of states (12a) and (12b).  Compare
with an observer who makes a local measurement that distinguishes
between states
$$
|in~D_1\rangle |in\rangle_{D_1} |r\rangle_{D_2}  \eqno(13a)
$$
and
$$
   |in~D_2\rangle |r\rangle_{D_1} |in\rangle_{D_2} . \eqno(13a)
   $$
   The records of the latter will be accurate even after the
   interaction with the environment.  Thus, the sentient being who
   thinks in terms of {\it local} properties has an evolutionary
   advantage due to the stability of local states (such as (13a) and
   (13b)).  An extensive research, led by Zurek (1993), of the role of
   the environment in the measuring process shows the stability of
   local events in the causal chain of a human observer.  I believe
   that the decomposition of the Universe into sensible worlds (11)
   is, essentially, unique.  The decomposition, clearly, might differ
   due to coarse or fine graining, but to have essentially different
   decompositions would mean having a multi-meaning Escher-type
   picture of the whole Universe continuously evolving in time.
   
   Recently Saunders (1993), equipped with heavy formalism of {\it
     decoherent histories} developed by Gell-Mann and Hartle (1990),
   investigated (in the framework of MWI) a model of ``evolutionary
   adaptation''. It has the above mentioned elements of locality and
   stability. Although I do not necessarily accept his model, I am
   certainly encouraged by his conclusion: ``These arguments [his
   arguments for evolutionary adaptation] are qualitative, but it
   seems that there is no difficulty in principle in construction of
   more detailed models.''

\vskip 1cm

 \centerline{\bf 14.  God Does Not Play Dice}
 \vskip .2cm 

The statement ``God does not play dice'' is probably the most famous
objection Einstein had to quantum theory.  The quantum theory with
collapse introduced a new type of probability, not an effective
probability due to our ignorance about exact details of the state
prior to a measurement, but a probability of genuinely unpredictable
outcomes.  Quantum events are such that even God (or infinitely
advanced technology) cannot predict them.  Bell (1964) proved that
unless God has some nonlocal features, which is in conflict with
Einstein's even more sacred  principle, God cannot predict the outcomes of
some quantum measurements performed on a simple system of two spin-1/2
particles.

The MWI solves the difficulty of the genuinely random Universe.  God
does not play dice. Everything is deterministic from the point of view
of God.  Everything evolves in time according to Schr\"odinger's
equation.  At the same time, there is an explanation of why for {\it
  us} there is genuine unpredictability when a quantum measurement is
performed. Ballentine (1975), however, claims that God does play dice
even in the framework of MWI. He plays dice when he assigns the ``me''
whom I know to a particular world.  However, at least in my version of
MWI, God does not and {\it cannot} do it.  I am in a privileged
position relative to an external observer, including God, in my
ability to identify myself without specifying the world in which I am.
God can identify me only by the world in which I am.  Therefore, God
cannot assign the ``me'' whom I know to a given world, and he cannot
define an objective {\it probability} for the ``me'' whom I know
ending up in a particular world.  Compare with the discussion of the
neutron with a sleeping pill and an experimenter in Section 7.

Very recently Page (1995)  supported this approach by arguing  that
``probabilism is a myth''. He developed a ``sensible quantum
mechanics'', a variant of MWI in a ``many-perceptions'' framework.
Although it has some resemblance with the many-minds interpretation of
Albert and Lower (1988), the latter is different: the
``minds'' do evolve probabilistically. See also a thorough
philosophical analysis of various alternatives by
Butterfield (1995).

The concept of probability in MWI is very different from our usual
probability.  Previously, we always used the concept of probability
when one of several possibilities would take place; but according to
MWI {\it all} these possibilities are realized in the Universe.  I
believe, however, that I have succeeded in introducing a concept of
subjective probability for sentient beings in each separate world,
while leaving the whole Universe deterministic.  The probability
postulate - probability is proportional to the measure of existence -
explains the only thing which, I think, requires an explanation: an
experimental fact about the consistency of frequencies of outcomes of
quantum measurements (performed in {\it our} world) with statistical
predictions of standard quantum theory.  Indeed, the sum of measures
of existence of all such worlds is overwhelmingly larger than the sum
of measures of existence of worlds in which the frequencies of the
quantum measurements differ significantly from those predicted by the
quantum theory.  (The latter worlds also exist, but in these worlds
sentient beings have no reason to believe that the quantum theory is
correct.)

Up to the present, there continues an extensive debate (see Kent, 1990
and references there) about the possibility, using MWI, of {\it
  deriving} the quantum probability rule from a weaker probability
postulate.  The claim is that from the postulate that the probability
of result $i$ is 0 when $|\alpha_i|^2 = 0$ and 1 when $|\alpha_i|^2 =
1$ (see Eq.  (11)) it {\it follows} that the probability for the
result $i$ is equal to $|\alpha_i|^2 $ for {\it any} value of
$\alpha_i$.  I agree with the opponents of MWI that the assumption of
the existence of many worlds does not help to derive the quantum law
of probability.  This debate has reflected badly on MWI. Failure of
MWI to be useful in deriving the quantum law of probability is
frequently - but wrongly - considered to be a proof of its inadequacy.
In fact, no other interpretation is better in this respect.
 
 \vskip 1cm

 \centerline{\bf 15.  The Causal Interpretation}
  \vskip .2cm
 
  As a physical theory, the MWI is more economical than any other
  quantum theory without collapse. The no collapse assumption
  invariably leads to the existence of the State of the Universe with
  all its ``branches'' corresponding to all innumerable worlds of MWI.
  So all the complexity of of MWI is there and, in addition, there is
  something else. Let me, however, touch here one other leading
  non-collapse interpretation.
  
  A very interesting non-collapse theory is the ``causal
  interpretation.'' The most credit for it should be given to Bohm
  (1952), however earlier de Broglie (1927) and later Bell (1981) also
  contributed to this beautiful picture. In addition to the quantum
  state of the Universe there is a {\it point} in configuration space
  of locations of all particles. The motion of this point is governed
  by the values of the wave function in the immediate vicinity (in
  configuration space) of the point according to a simple equation
  (especially simple in Bell's version of the theory).
  
  Proponents of the causal interpretation frequently consider both the
  wave function and the Bohmian particle (the point) as ``physically
  real''. I, however, find the most fruitful approach to the Bohm
  theory the interpretation according to which only {\it the point}
  corresponds to ``reality'', while the wave function is a secondary
  entity, whose purpose is to be a ``pilot'' of the point. Only in
  this way I can see how the causal interpretation describes a single
  ``real'' world.
  
  The paradoxes of non-relativistic quantum mechanics are explained
  beautifully by the causal interpretation. The particles, after all,
  do have trajectories. The neutron (the corresponding coordinate of
  the point) in the neutron interference experiment passes through one
  of possible trajectories, while the corresponding wave passes
  through both of them. The theory is also deterministic from the
  point of view of God, so he does not play dice. (The argument of
  Bell for the unpredictability of spin measurements performed on a
  certain pair of spin-1/2 particles is solved explicitly according to
  the Bohr's vision: the outcome of a spin measurement is certain only
  when we specify the measuring device we use.)
  
  So it seems that the causal interpretation has all the good
  properties of MWI and it has only a single world: surely a desirable
  feature. I, however, do not consider it preferable to MWI. Besides
  the main technical problem, the lack of reasonable generalization to
  relativistic domain I want to mention two other points.
  
  The first point is a peculiar and, in my opinion, unfortunate
  feature of the causal interpretation which has been understood only
  recently.  Bell (1980) pointed out that in some situations the Bohm
  trajectory is very different from what we would naively think, and
  moreover, Englert {\it et al.} (1992), Brown {\it et al.} (1995),
  and Aharonov and Vaidman (1996) have shown that it might be
  different from what an apparently good measuring device would show.
  Consider the experiment described in Figure 4. A single neutron
  entered the beam splitter $S_1$ and has been detected by, say,
  detector $D_1$. Then, we would say that its trajectory is $S_1 M_2
  D_1$. But the Bohm trajectory (called by Englert {\it et al.}(1992)
  ``surrealistic trajectory'') in this case is $S_1 M_1 A D_1$. The
  neutron changes the direction of its motion at point $A$ in spite of
  the fact that no beam splitter is there. Moreover, if we try to
  observe the trajectory of the neutron we might be ``fooled'' by our
  measuring apparatus.  Let us assume that the experiment is performed
  in a special ``bubble chamber'', such that the mechanism of creation
  a bubble by the neutron is such that the neutron changes an internal
  state of some atom and then the atom slowly creates the bubble. This
  process is slow enough that during the time the neutron passes from
  the beam splitter $S_1$ till the time it reaches the intersection
  point of the two possible trajectories $A$ (see Figure 5) the wave
  function of the excited atom does not change significantly its
  spatial distribution. In this case, the Bohm trajectory is again
  $S_1 M_1 A D_1$ while the bubbles, developed after the neutron
  passage, will show (naively expected) $S_1 M_2 D_1$.

\epsfysize=8.5 cm
 \centerline{\epsfbox{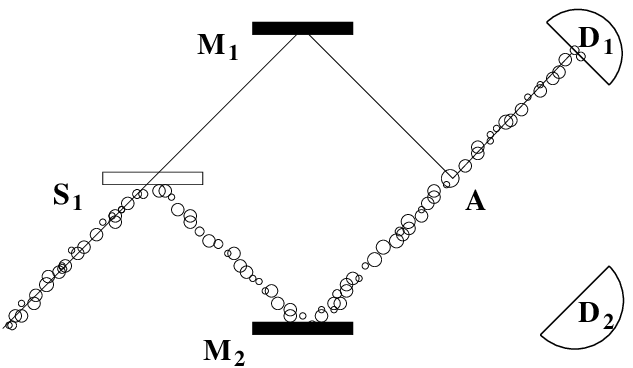}} 
{\small {\bf Figure 5}: {\footnotesize 
The neutron interferometer without second beam  splitter in a bubble chamber.
}}
\vskip .2cm

The last point is probably the most important argument against
accepting the causal interpretation or any other non-collapse
interpretation which has some additional elements.  I do not see that
these interpretations really get rid of all but one world. If a
component of the quantum state of the Universe, which is a wave
function in a shape of a man, continues to move (to live?!) exactly as
a man does, in what sense it is not a man? How do {\it I} know that I
am not this ``empty'' wave?
 
\vskip .8cm

\centerline{\bf 16.  Why MWI?}
\vskip .2cm 

The crucial argument in favor of MWI is that in this theory there is
no collapse to be explained.  The bad features of the collapse cannot
be overestimated.  ``The reduction [collapse] postulate is an ugly
scar on what would be a beautiful theory if it could be removed",
Gottfried's phrase (1989), represents the feeling of many physicists.
There is no clue to when exactly the collapse occurs.  If it does
occur, it seems impossible to avoid contradictions with special
relativity.  In spite of persistent efforts in the last half century,
there is no satisfactory physical explanation of the collapse.
Extremely divergent proposals for the cause of the collapse, such as
consciousness (Wigner, 1962), gravitation (Penrose, 1994, pp.335-347),
new genuine random processes (Ghirardi and Pearle, 1990), etc.
indicate the difficulties in the task of explaining the collapse.

For me, an important positive feature of MWI is the
elimination of conceptually unpredictable outcomes from the
fundamental theory of the Universe (God does {\it not} play dice).  I
want to believe, that at least in principle, Science can explain
everything.

Most physicists who favor MWI do so because it allows them to consider
the quantum state of the Universe, the basic concept in quantum
cosmology (e.g. Clarke, 1974).  The standard approach requires an {\it
external observer} for a system in a quantum state and, therefore, is
unable to deal with the quantum state of the whole Universe.

Although nobody has done it explicitly, it seems that MWI can be
extended to the relativistic domain because all paradoxes of
superluminal changes disappear with the removal of the collapse. For
discussion of quantum nonlocality in the framework of MWI see Vaidman
(1994).

MWI yields a novel basis for the investigation of the relation between
mind and matter.  According to MWI, a human being
{\it is} a wave function which is a part of a quantum state which
{\it is} the world, which in turn is one term in the superposition
of many quantum states which comprise the State, which {\it is} the
Universe.

It might be worthwhile to make an attempt to learn about the ``other''
worlds by investigating records of quantum measurement-type
interactions.  Thus we will obtain some information about the whole
Universe, beyond our subjective world.  This information might lead to
better understanding of the problems of evolution.

MWI can greatly influence the ongoing discussion (see Healey, 1992) of
causation, EPR correlations etc.  It provides a deterministic physical
theory without nonlocal interactions.

Although one does not have to believe in MWI in order to design a
machine which employs quantum interference on a macroscopic scale, it
is clearly more natural to discuss these possibilities when one does
not need to worry about ``miraculous'' collapses, but only about
quantum correlations described by Schr\"odinger's equation.  It is not
a coincidence that the pioneer of ``quantum parallel processing'' is
an enthusiastic proponent of MWI -- Deutsch (1985).  While it is
hopeless to reach the ``other'' worlds which are already split from
``our'' world, it is feasible to create several worlds carefully and
to reunite them later. This is, essentially, the subject of current
intensive research of building a {\it quantum computer} which splits
to make many different calculations in parallel and reunites to give
the final result. Recently Shor (1994) has shown that quantum computer
can solve some important problems significantly faster than any
existing algorithm of classical computation.

I have one more reason to be enthusiastic about MWI (see Vaidman,
1994). It helps me to see and understand novel features of quantum
mechanics.  Thinking in terms of MWI was especially fruitful in recent
work by Elitzur and myself (1993).

\vskip .8cm

 \centerline{\bf 17. Reflections on the `Many Minds' Interpretation}
\vskip .2cm
  
While I was correcting the manuscript according to suggestions of the
referees British Journal for the Philosophy of Science has published
an enlightening Symposium on `Many Minds' Interpretations of Quantum
Mechanics. The issues discussed in the Symposium are very closely
related to what I have described above and I believe that my brief
reflections on the Symposium will help to understand my MWI and will
avoid misunderstandings due to different terminology.  

My MWI is much more close to the Many Minds Interpretation (MMI) of
Lockwood (1996) than to the MWI he refers to. The latter seems to be
inspired by De Witt (1970) and I have criticized it myself (see Sec.
10). Papineau's (1996) illuminating analysis of ``what it {\it would
  be like} to have a superposed brain'' in the framework of MMI
corresponds exactly to my understanding of MWI; and I am ready to sign
under Lockwood sentence (1996, p.170): ``A many minds theory, as I
understand it, is a theory which takes completely at face value the
account which unitary quantum mechanics gives of the physical world
and its evolution over time'' with the only change of ``many minds''
to ``many worlds''.

 Indeed, I believe that the name ``Many Worlds'' is more
appropriate; I defend it in Section 1. I also find support to this
choice in the Symposium:
Deutsch (1996), apart from analyzing strong (although indirect)
evidences for existence of many worlds (which he names ``universes''),
gives personal testimony for Everett's similar view.

Beyond semantics, I do not think that Lockwood's concept of ``Mind''
is as important and fruitful as he suggests. What corresponds to our
everyday experience is Lockwood's ``mind''. The ``Mind'' is the
concept which is relevant only for ``god'' (infinitely advanced
technology). But even god might find this concept not very clear
because of the ambiguity related to the birth of Mind: the mother
might be in a superposition of being pregnant for different times and
from different fathers.
 
Another argument in favor of ``world'' relative to ``mind'' (or
``Mind'') is that in the many worlds picture we are not forced to
accept ``radicalism'' discussed by Butterfield (1996) according to
which ``unobserved macroscopic world can be very indefinite, even
within a branch''. In my MWI a world is a ``sensible story''. Thus, if
we are not ready to consider stars (even those we cannot see now)
being in a superposition of states corresponding to macroscopically
different locations, we can always choose an appropriate basis such that
we will have stable classical-type stories. We also might discuss many
worlds at far past (or far future?!) when no conscious minds were
(will be) present by choosing the basis according to understanding
pattern of present sentient beings.

However, the main aspect of ``mind'' -- its role for defining {\it
  preferred basis} -- I see exactly as Lockwood does.  As it is
stressed in the Symposium by Lockwood, Papineau, Saunders (1996) and
others, the preferred basis is not fixed by fundamental physics. The
basis is defined by a sentient observer. The fundamental physics,
however, due to locality of the basic interactions leads, via {\it
  decoherence}, to existence of only certain types of sentient beings.
I cannot see any difficulty with this explanation, but in the
Symposium I found only cautious appeal to decoherence, see Brown
(1996).

Butterfield (1996, p.203) points out that in MWI the preferred basis
need not be fixed once and for all and he argues that this is a
disadvantage relative, say, to the Bohm theory. I agree that a theory
which gives the basis (and even better, a specific choice of the basis
vector) is preferable. However, it seems to me that the price for this
definiteness is too high and it is not clear that the goal is really
achieved: Deutsch (1996), in his discussion of ``unoccupied grooves''
or `mindless hulk', has essentially the same position as I do viewing
the Bohm theory to be a many-worlds theory.

Probably the closest point between my MWI and Lockwood's MMI is the
issue of probability which got a lot of attention in the Symposium.
Lockwood (1996, p.182) insists on ``the existence of a naturally
preferred {\it measure} ...'' which is essentially ``the measure of
existence of a world'' which I introduce. Papineau (1996) reinforces
the introduction of such a measure in MMI (MWI) showing that the
concept of probability is in no way better if not worse in other
theories.  Lower and Butterfield, however, argue that such a
probabilistic measure requires introducing ``persisting minds'' which
I do not have in my MWI (except for connection through common memories
which might yield an answer to Butterfield).  Lower (1996, p.230)
writes: ``it doesn't make any sense on the Instantaneous Mind View
speak of the probability of an instantaneous mind at $t$ evolving to
exemplify a mental state $M$ at $t'$ since there is no fact of the
matter concerning the transtemporal identity of minds''.  It seems to
me that this is exactly the problem I raised and solved in section 7.
Lower realized that the key issue for defining the probability measure
is ``personal identity''.  The probability of the mind at $t$ to
evolve to the mental state $M$ at time $t'$ is defined as an ignorance
probability of the ancestors of the mind at (or prior to) time $t'$ to
be in the state $M$.  In fact, I believe that this and the gedanken
experiment showing physical significance of the measure of existence
of the present world (Section 11) are the main novel points I made
here. To summarize, in this paper I went beyond mathematical
definition of the probability measure: I succeeded to attach the
ignorance-probability meaning to the measure of existence of the
future worlds and I found a physical meaning for the measure of
existence of the present world by designing a gedanken situation in
which one should behave differently only because her world has
different measure of existence.

\vskip 0.1 cm 

\noindent
{\it Acknowledgments} -- I am grateful to many friends, colleagues and
referees for their patience in the endless discussions which resulted
in this paper, and especially to the late Ferdy Schoeman to whom I
dedicate this work.  This research was supported in part by grant
614/95 of the Basic Research Foundation (administered by the Israel
Academy of Sciences and Humanities).

\vskip .8cm

 \centerline{\bf References}
\vskip .15cm
\footnotesize

\vskip .13cm \noindent Aharonov, Y. and Vaidman L. (1996) `About
Position Measurements Which Do Not Show the Bohmian Particle
Position', in J. T.  Cushing, A. Fine, and S. Goldstein (eds.), {\it
  Bohmian Mechanics and Quantum Theory: an Appraisal} (Netherlands:
Kluwer Academic Publishers) pp. 141-154.

\vskip .13cm \noindent 
Albert, D. (1987)
 `A Quantum-Mechanical Automaton', {\it Philosophy of Science}
{\bf 54}, 577-585.

\vskip .13cm \noindent 
 Albert D. and B. Lower, B. (1988)
 `Interpreting the Many Worlds Interpretation', {\it
Synthese} {\bf 77}, 195-213.

\vskip .13cm \noindent 
Ballentine, L. E. (1975)
 {\it Measurements and Time Reversal in Objective Quantum
Theory} (Oxford: Pergamon Press), pp. 50-51.

\vskip .13cm \noindent 
Bell, J. S. (1964)
 `On the Einstein Podolsky Rosen Paradox', {\it
  Physics} {\bf 1},  195-200.

\vskip .13cm \noindent 
 Bell, J. S. (1981)
  `Quantum Mechanics for Cosmologists' in C.J.  Isham, R. Penrose
and D.W. Sciama (eds), {\it Quantum Gravity 2: A Second Oxford Symposium}
(Oxford: Caledonia Press) pp. 611-637.

\vskip .13cm \noindent 
 Bell, J. S. (1980)
 `de Broglie-Bohm, Delayed-Choice Double-Slit Experiment,
and Density Matrix', {\it International Journal of Quantum Chemistry} {\bf
14}, 155-159.
 
\vskip .13cm \noindent 
Ben-Dov, Y. (1990)
  `An ``Observer Decomposition'' for Everett's Theory', {\it
Foundation of Physics Letters} {\bf 4}, 383-387.

\vskip .13cm \noindent 
Bohm, D. (1952) 
``A Suggested Interpretation of the Quantum Theory in Terms of
`Hidden' Variables I and II'', {\it Physical Review} {\bf 85}, 97-117.

\vskip .13cm \noindent 
Brown, H. R., Dewdney C. and Horton ? (1995)
`Bohm particles and
their detection in the light of neutron interferometry', {\it
  Foundations of Physics} {\bf  25}, 329-347.

\vskip .13cm \noindent 
Brown, H. R. (1996)
`Mindful of Quantum Possibilities',
 {\it British Journal for the Philosophy of Science} {\bf 47}, 189-199.

\vskip .13cm \noindent 
Butterfield, J. (1995)
 `Worlds, Minds and Quanta', in {\it Aristotelian
  Society Supplementary Volume}, {\bf 69}, pp.113-158.

\vskip .13cm \noindent 
Butterfield, J. (1996)
`Whither the Minds',
 {\it British Journal for the Philosophy of Science} {\bf 47}, 200-221.

\vskip .13cm \noindent 
Clarke, C. J. S.  (1974)
 `Quantum Theory and Cosmology', {\it Philosophy of Science}
{\bf 38}, 317-332.

\vskip .13cm \noindent 
De Broglie, L. (1927)
 `La Mechanique Ondulatoire et al Structure Atomique de la
Matiere et du Rayonnement', {\it Journal de Physique et le Radium, Serie
VI} {\bf 8}, 225-241.

\vskip .13cm \noindent Deutsch, D. (1985) `Quantum Theory, the
Church-Turing Hypothesis, and Universal Quantum Computers', {\it
  Proceedings of Royal Society of London} {\bf 400}, 97-117.

\vskip .13cm \noindent 
Deutsch, D. (1986)
 `Three Connections Between Everett's Interpretation and
Experiment', in R. Penrose and C.J. Isham (eds), {\it Quantum Concepts of
Space and Time} (Oxford: Caledonia Press, 1986), pp. 215-226.

\vskip .13cm \noindent 
Deutsch D. (1996)
`Comment on Lockwood',
 {\it British Journal for the Philosophy of Science} {\bf 47}, 222-228.

\vskip .13cm \noindent 
De Witt, B. S. (1970)
 `Quantum Mechanics and Reality' {\it Physics Today} {\bf
23}, 30-35.

\vskip .13cm \noindent Elitzur, A. and Vaidman, L. (1993)
`Interaction-Free Quantum Measurements', {\it Foundation of Physics}
{\bf 23}, 987-997.

\vskip .13cm \noindent 
Englert, B., Scully, M. O., S\"ussmann, G. and Walther, H. (1992)
`Surrealistic Bohm  trajectories',
 Zeitschrifft f\"ur Naturforschung 47a, 1175-1186.

\vskip .13cm \noindent 
Everett, H. (1957)
 `{\it Relative State}
  Formulation of Quantum Mechanics', {\it Review of Modern Physics},
  {\bf 29}, 454-462.

\vskip .13cm \noindent 
Gell-Mann M.  and Hartle, J. B. (1990) 
`Quantum Mechanics in the Light of Quantum Cosmology',
 in W. H. Zurek (ed), {\it Complexity, Entropy  and the Physics 
of Information}, (Reading: Eddison-Wesley), pp. 425-459.

\vskip .13cm \noindent 
Ghirardi G. C. and Pearle, P. (1990)
 `State Vector reduction I and II,
Elements of Physical Reality, Nonlocality and Stochasticity in
Relativistic Dynamical Reduction Models', in A. Fine and M. Forbes (eds), {\it Philosophy of Science
  Association 1990}, Volume II, (East
Lancing: PSA), pp. 19-47.

\vskip .13cm \noindent 
Gottfried, K. (1989)
 `Does Quantum Mechanics Describe the ``Collapse"
of the Wave Function?',  Contribution to the Erice School:{\it 62 Years
of Uncertainty}, unpublished.

\vskip .13cm \noindent 
Greenberger, D. M. (1983)
 `The Neutron Interferometer as a Device for
Illustrating the Strange Behavior of Quantum Systems', {\it Reviews of
  Modern Physics} {\bf 55}, 875-905.

\vskip .13cm \noindent 
Healey, R. A. (1984)
 `How Many Worlds?' {\it Nous}  {\bf 18},  591-616.

\vskip .13cm \noindent 
Healey, R. A. (1992)
 `Discussion: Causation, Robustness and EPR',
{\it Philosophy of Science}  {\bf 59}, 282-292.

\vskip .13cm \noindent 
Horne, M.A., Shimony, A.  and Zeilinger, A. (1989)
`Two-Particle Interferometry', {\it Physical Review Letters} {\bf 62},
2209-2212.

\vskip .13cm \noindent 
Kent, A. (1990)
 `Against Many-Worlds Interpretation', {\it International
Journal of Modern Physics} {\bf A5}, 1745-1762.

\vskip .13cm \noindent 
Lewis, D. (1986)
  {\it On the Plurality of Worlds} (Oxford, New York:
Basil Blackwell).

\vskip .13cm \noindent 
Lockwood, M. (1989)
 {\it Mind, Brain {\rm \&} the Quantum} (Oxford, and
Cambridge, MA: Basil Blackwell).

\vskip .13cm \noindent 
Lockwood, M. (1996)
```Many Minds' Interpretations of Quantum Mechanics',
 {\it British Journal for the Philosophy of Science} {\bf 47}, 159-188.

\vskip .13cm \noindent 
Lower, B. (1996)
`Comment on Lockwood',
  {\it British Journal for the Philosophy of Science} {\bf 47}, 229-232.

\vskip .13cm \noindent 
Page, D. (1995)
 `Sensible Quantum Mechanics: Are Only Perceptions Probabilistic?'
{\it University of Alberta report}:  quant-ph/9507006. 

\vskip .13cm \noindent 
Papineau, D. (1996)
`Many Minds No Worse than One',
 {\it British Journal for the Philosophy of Science} {\bf 47}, 233-340.

\vskip .13cm \noindent 
Penrose, R. (1994)
 {\it Shadows of the Mind} (Oxford: Oxford
University Press).

\vskip .13cm \noindent 
Redhead M. and Teller, P. (1992)
`Particle Labels and the Theory of Indistinguishable Particles in
Quantum Mechanics', {\it British Journal for the Philosophy of
  Science} {\bf 43}, 201-218.

\vskip .13cm \noindent 
Saunders, S. (1993)
 `Decoherence, Relative States, and Evolutionary
Adaptation', {\it Foundations of Physics} {\bf 23}, 1553-1585.

\vskip .13cm \noindent 
Saunders, S. (1996)
`Comment on Lockwood',
  {\it British Journal for the Philosophy of Science} {\bf 47}, 241-248.
  
  \vskip .13cm \noindent Shor P. (1994) `Algorithms for Quantum
  Computation: Discrete Logarithms and Factoring', in S. Goldwasser
  (ed), {\it Proceedings of the 35th Annual Symposium on the
    Foundations of Computer Science}, (Los Alamitos: IEEE Computer
  Science Press) pp. 124-134.

\vskip .13cm \noindent 
Skyrms, B. (1976)
 `Possible Worlds, Physics and Metaphysics',
{\it Philosophical Studies} {\bf 30},  323-32.

\vskip .13cm \noindent 
Vaidman, L. (1994)
`On the Paradoxical Aspects of New Quantum Experiments'
 {\it Philosophy of Science Association 1994}, pp. 211-217.

\vskip .13cm \noindent 
Wigner, E.P. (1962)
 `Remarks on the Mind-Body Problem', in I. J. Good (ed), {\it
The Scientist Speculates} (London: Heinemann), pp. 284-302.

\vskip .13cm \noindent 
Zurek, W. H. (1994)
 `Preferred States, Predictability, Classicality,
and the Environment-Induced Decoherence', in J. J.  Halliwell, J.
Perez-Merzader and W. H.  Zurek, (eds), {\it Physical Origins of Time
 Asymmetry}, (Cambridge: Cambridge University Press).

\end{document}